\documentclass[twocolumn,byrevtex,prl,floatfix,showpacs,a4paper]{revtex4}
\usepackage{amssymb,amsmath,bm}
\usepackage{graphics}

\newcommand*{\br}{\ensuremath{\bm{r}}}

\begin{document}

\title{\bf {Equilibrium swelling and universal ratios in dilute
polymer solutions: Exact Brownian dynamics simulations for a delta
function excluded volume potential }}

\author{K.~Satheesh~Kumar}
\author{J.~Ravi~Prakash}
\affiliation{ Department of Chemical Engineering, Monash
University, Clayton, Victoria\textendash 3168, Australia}
\email{ravi.jagadeeshan@eng.monash.edu.au}
\date{\today}

\begin{abstract}

A narrow Gaussian excluded volume potential, which tends to a
$\delta$-function repulsive potential in the limit of a
\textit{width} parameter $d^*$ going to zero, has been used to
examine the universal consequences of excluded volume interactions
on the equilibrium and linear viscoelastic properties of dilute
polymer solutions. Brownian dynamics simulations data, acquired
for chains of finite length, has been extrapolated to the limit of
infinite chain length to obtain model independent predictions. The
success of the method in predicting well known aspects of static
solution properties suggests that it can be used as a systematic
means by which the influence of solvent quality on both
equilibrium and non-equilibrium properties can be studied.

\end{abstract}


\maketitle

There is a growing recognition of the importance of solvent
quality in determining the rheological properties of dilute
polymer solutions. Though there are relatively few
systematic experimental investigations of the influence of solvent
quality, nevertheless there is sufficient evidence that material
functions, in both shear and extensional flows, are significantly
different from each other in good and
$\theta$-solvents~\cite{nodetal,solmul96,sridharetal00}.
While the theoretical description of the influence of solvent
quality on the \textit{equilibrium} behavior of dilute polymer
solutions has been a major area of research~\cite{schafer}, the
description of its influence on \textit{non-equilibrium} behavior
is still in its infancy. Most current non-equilibrium
theories do not attempt to develop a unified framework that is
also applicable at equilibrium, and the wealth of experience and
insight that has been gained so far in the development of static
theories is rarely used in the description of dynamic behavior.
In this paper, we introduce an approach that provides a consistent
means of describing both regimes of behavior, and as a first step
towards establishing the usefulness of the methodology, we
demonstrate that the method is capable of reproducing well known
universal static results. We also show that, even in the case of
well established results, the method is capable of providing new
and additional insights. Furthermore, as an example of the
versatility of the approach, the influence of solvent quality on a
universal ratio of \textit{linear viscoelastic} properties, is
reported here for the first time.

One of the most important results of the experimental
investigation of the static behavior of dilute polymer solutions,
has been the discovery that various properties---both in
$\theta$-solvents and in good solvents---exhibit power law
behavior when the molecular weight of the dissolved polymer is
sufficiently large. It is perhaps less commonly known that even when the
molecular weight is not very large, a type of scaling still
persists, which enables the description of behavior with the help
of a single parameter. An illustrative example is the behavior of
the root mean square radius of gyration, $R_\text{g}$. Light
scattering measurements indicate that for molecules of
sufficiently large molecular weight $M$, $R_\text{g}$ obeys the
power law $R_\text{g} = a M^\nu$, where the pre-factor $a$ depends
on the particular polymer-solvent system, but the exponent $\nu$
does not. In a $\theta$-solvent, the \textit{universal} exponent
has a value, $\nu = 0.5$, while in a good solvent, $\nu = 0.592
\pm 0.003$~\cite{haygra99}. A remarkable experimental observation
is that, away from the large molecular weight asymptotic limit,
the dependence of the mean size of the polymer molecule on both
the temperature, $T$, of the solution and $M$,
can be combined into a single variable, $\tau(T)
\sqrt M$, where, $\tau(T)$ is a simple function of temperature,
$\tau = (1-{T_\theta}/{T})$, with $T_\theta$ denoting the
$\theta$-temperature. A vast amount of equilibrium data for a
variety of polymer solvent systems reveals that, when the
\textit{swelling} $\alpha_\text{g}$ of a polymer
molecule---defined as the ratio of the radius of gyration in a
good solvent to the radius of gyration in a $\theta$-solvent---is
plotted as a function of $\tau \sqrt M $ (with each
polymer-solvent system being shifted horizontally by a suitable
constant factor), universal behavior is displayed, for all values
of the molecular weight $M$, and temperature $ T > T_\theta
$~\cite{miyaki81,haygra99,berceaetal99}.

The entire range of behavior exhibited by static solution
properties has been successfully predicted by
applying renormalization group (RG) methods~\cite{schafer}.
In these theories, the macromolecule is frequently represented by a
\textit{coarse-grained} model, such as a bead-spring chain,
consisting of $N$ beads connected together by $(N-1)$
\textit{Hookean} springs (with a spring constant $H$). The
presence of excluded volume interactions is then taken into
account by assuming the existence of a Dirac delta function
repulsive potential, that acts pairwise between the beads of the
chain~\cite{doiedw},
\begin{equation}
E \left( {\br}_{\nu \mu} \right) = v (T) \, k_{\rm B} T \, \delta
\left( {\br}_{\nu \mu} \right) \label{deltapot}
\end{equation}
where, $v (T) $ is the excluded volume parameter,
$ k_{\rm B}$ is Boltzmann's constant, and ${\br}_{\nu
\mu}$ is the vector connecting beads $\nu$ and $\mu$. In the limit
$T \to T_\theta$, it can be shown that $v$ depends on $T$ through
the relation, $v(T)= v_0 \, \tau (T)$, where $v_0$ is a
constant~\cite{doiedw}. If a length scale, $\ell =
\sqrt{k_{\text{B}} T /H }$, and a non-dimensional
\textit{strength} of excluded volume interactions, $z^* = v \left(
2 \pi \ell^2 \right)^{- 3 / 2}$ are defined, then both the
temperature and the chain length dependence can be combined into
the single non-dimensional parameter, $z=z^* \sqrt N$. Completely
consistent with experimental observations (since $z \propto \tau
\sqrt M $), RG theories predict the existence of power laws in the
limit $z \to 0$ (corresponding to $\theta$-solvents), and the
excluded volume limit $z \to \infty$ (corresponding to good
solvents), and the existence of \textit{scaling functions}, that
depend only on the parameter $z$, which accurately describe the
crossover behavior between these two asymptotic
limits~\cite{schafer}.

In spite of the success of RG theories, it must
be borne in mind that they are approximate theories, dependent to
some extent on the order of the original perturbation calculation
on which they are based. Exact numerical results for the excluded
volume problem can, on the other hand, be obtained by either Monte
Carlo simulations of self avoiding walks on
lattices~\cite{lisokal95,grassberger97}, or by off-lattice Monte
Carlo simulations~\cite{greetal99}. Since $\delta$-functions
cannot be used in numerical investigations, off-lattice Monte
Carlo simulations are typically based on excluded volume
potentials with a finite range of excluded volume interactions,
such as the Lennard-Jones potential. An important point to be
noted, however, is that the parameter $z$, which is the true
measure of solvent quality, does not appear naturally when
potentials like the Lennard-Jones potentials are used. Instead,
simulation results are functions of the parameters that
characterize the potential. It is possible,
however, to infer the value of $z$ from simulation results by
carrying out a data shifting procedure similar to that used to
bring experimental data onto a universal curve, i.e, data for
various values of chain length $N$, and various distances from the
$\theta$-temperature, can be brought to lie on a master
curve~\cite{greetal99}. The procedure is quite involved and even
the $\theta$-temperature must first be estimated from simulations.
Furthermore, Monte Carlo methods cannot be used to estimate
rheological properties in general flow fields, and so it is not
clear how this procedure can be extended to systematically examine
the influence of solvent quality on non-equilibrium properties.

The self-similar character of polymer molecules, which is
responsible for the universal behavior exhibited by polymer
solutions at equilibrium, is also responsible for the universal
behavior displayed away from equilibrium~\cite{doiedw}.
An important and fundamental challenge for molecular theories
is to verify the existence of this universal behavior. The
universal consequences of excluded volume effects away from
equilibrium has been examined by \"Ottinger and
co-workers~\cite{ott89c,zylott91} with the help of RG methods.
Since these methods lead to approximate predictions, it is
important to be able to assess their accuracy. Further, these
papers were focussed on predicting the universal behavior in the
excluded volume limit, and cross-over scaling functions for
rheological properties were not reported.

In this paper, we introduce a procedure by which universal
cross-over scaling functions and asymptotic behavior in the
excluded volume limit can be obtained for a $\delta$-function
excluded volume potential, both at equilibrium and away from
it. In other words, a means has been obtained by
which an \textit{exact} numerical solution to
the Edwards model, which is of fundamental importance to polymer
physics, can be found. Using the suggested methodology, provided sufficient
computational time is used, properties predicted by the
Edwards model can be found to a high degree of accuracy.
As a result, the influence of solvent quality,
measured directly in terms of the parameter $z$,
can be systematically examined, and results obtained by approximate
RG calculations may be verified.

A detailed discussion of the theoretical framework used here
may be found in earlier papers by~\textcite{pra01a,pra02}.
Essentially, macroscopic properties are obtained by
exploiting the mathematical equivalence of diffusion equations in
polymer configuration space and stochastic differential equations
for the polymer configuration. Since averages calculated from
stochastic trajectories are identical to averages calculated from
distribution functions, stochastic trajectories are generated by
numerically integrating the appropriate stochastic differential
equation with the help of Brownian dynamics simulations
(BDS)~\cite{ottbk}.

A key ingredient in our approach is the use of a
narrow Gaussian potential (NGP),
\begin{equation}
E \left( {\br}_{\nu \mu} \right) = \left( \frac{z^* }{
{d^*}^3} \right) k_{\text{B}} T \, \exp \left\lbrace - \frac{H }{
2 k_{\text{B}} T }\, \frac{ \br_{\nu \mu}^2 }{ {d^*}^2}
\right\rbrace \label{evpot}
\end{equation}
where, $d^*$ is an additional non-dimensional parameter that
measures the range of excluded volume interaction. The NGP
is a means of regularizing the
$\delta$-function potential since it reduces to the
$\delta$-potential in the limit of $d^* \to 0$. The central
hypothesis behind the use of this potential is that when data
accumulated for chains of finite length is extrapolated to the
infinite chain length limit, universal consequences of the
presence of excluded volume interactions will be revealed,
independent of the value of $d^*$. Initial attempts to carry out
this program were unsuccessful because of the computational
intensity of BDS. The essential soundness of the procedure has,
however, been established by obtaining universal predictions,
both at equilibrium and in the linear viscoelastic
limit~\cite{pra01b}, and at finite shear rates in steady simple
shear flow~\cite{pra02}, with an approximate version of the
theory---the so called \textit{Gaussian
approximation}. The parameter $d^*$ has been shown
to be irrelevant in the long chain
limit, since it always appears in the
theory as the ratio $d^*/\sqrt{N}$.

An exact numerical solution of the model has been obtained in this
work, and the computational difficulties encountered earlier have been
overcome, by adopting two new strategies (apart from
the obvious one of parallelizing the code). First, a novel
trajectory in the ($d^*,z^*$) parameter space has been used to reach
the asymptotic limit. Second, a variance reduction procedure based
on the method of control variates~\cite{ottbk}, with the Rouse
model as the control variate, has been implemented. While these
strategies, which are discussed in greater detail
elsewhere~\cite{satpra02}, do not alter the scaling of CPU time
with $N$ (which continues to increase roughly as $N^3$), they
ensure that the noise is sufficiently small to permit an accurate
extrapolation of finite chain data, even for fairly small chains.

Two different approaches have been adopted here to obtain results
in the limit $N \to \infty$. In the first, $z$ is kept constant
while simulations are carried out with increasing values of $N$.
Extrapolation of the accumulated data to $N \to \infty$ implies
the simultaneous limit, $z^* \to 0$ (or equivalently, $T \to
T_\theta$). As a result, crossover functions for systems close to
the $\theta$-temperature are obtained. In the second approach,
$z^*$ is maintained constant (which is equivalent to $T$ being
constant) while $N$ is increased systematically. Extrapolation to
$N \to \infty$ in this case leads to the prediction of behavior in
the excluded volume limit, $z \to \infty$.
\begin{figure}[tb]
\resizebox{8.5cm}{!}{\includegraphics*{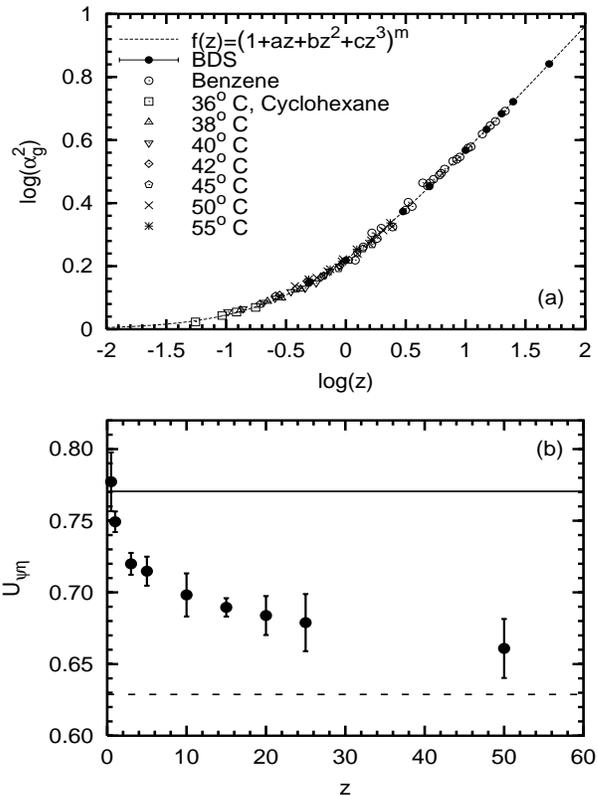} }
\caption{Crossover scaling functions. (a) Swelling of the radius
of gyration, $\alpha^2_\text{g}$, versus the strength of excluded
volume interaction, $z= z^* (T) \sqrt{N}$. Experimental data for
polystyrene in the two solvents is that of~\textcite{miyaki81}.
(b) Universal ratio $U_{\Psi \eta }$ as a function of $z$.
The solid line is the prediction in the excluded volume limit,
while the dashed line is the prediction of RG theory.
\label{fig1}}
\end{figure}

The crossover scaling function for the swelling
$\alpha^2_\text{g}$, as a function of $z$, is displayed in
Fig.~\ref{fig1}(a). The experimental data of~\textcite{miyaki81}
for polystyrene of various molecular weights and temperatures has
been shifted to lie on the simulation data, using a single
constant horizontal shift factor for each of the two
solvents~\cite{satpra02}. The remarkable agreement of
the predicted equilibrium swelling with experimental data,
similar to that obtained earlier by RG theory~\cite{schafer},
has been obtained by using bead-spring chains with a maximum of
just 36 beads. The BDS data has been
curve-fitted with an equation commonly used to fit results of RG
analysis~\cite{schafer}, $ \alpha^2_\text{g} = \left( 1 + a \, z +
b \, z^2 + c \, z^3 \right)^{(4 \nu -2)/3}$, with the coefficient
$a$ kept fixed at its first order perturbation
value~\cite{schafer}. It leads to the prediction $\nu = 0.6004 \pm
0.0004$, which is about 2\% greater than the value $\nu = 0.588$
predicted by RG theory. The maximum difference between the
computed data and the curve fit was less than 0.6\%. As shown
below, we find that $\nu$ estimated in the excluded volume limit,
with the present procedure, is also about 2\% more than 0.588.

The efficacy of the present approach is demonstrated in
Fig.~\ref{fig1}(b), where the crossover behavior of the
universal ratio constructed from the zero shear rate
first normal stress difference coefficient and the
zero shear rate viscosity, $U_{\Psi \eta }= {{n_\text{p}
k_\text{B} T \Psi_{1, 0} / { \eta_{p, 0}^2}}}$ (with $n_\text{p}$
denoting the number density of polymers), is displayed.
To our knowledge, this crossover behavior has not
been reported elsewhere before. The only existing
universal prediction is a RG calculation in the
excluded volume limit (indicated in the figure by
the dashed line), obtained by refining a first
order perturbation expansion~\cite{ott89c}.

In the Rouse model, $U_{\Psi \eta } = \textrm{constant} = 0.8$. It
is also expected to be constant, independent of the
polymer-solvent system, in the excluded volume limit. The figure
suggests that the crossover function approaches the value
predicted by RG theory, $U_{\Psi \eta } = 0.6288$, as $z \to
\infty$. On the other hand, as indicated by the solid line, a
significantly different value of $U_{\Psi \eta } = 0.77 \pm 0.01$
is predicted by the present approach in the excluded volume limit.
It is of interest to see if a RG calculation based on a higher
order perturbation theory would lead to a prediction in agreement
with the present prediction. A similar difference, between the
large $z$ limit suggested by the crossover behavior, and the value
obtained in the excluded volume limit, has been observed
previously in the case of the universal ratio $U_{R} =
\alpha^2_\text{g}/ \alpha^2_\text{e}$, where $\alpha^2_\text{e}$
is the swelling of the end-to-end vector. The behavior of $U_{R}$
predicted by the present approach will be discussed
elsewhere~\cite{satpra02}.

\begin{figure}[t]
\resizebox{8.5cm}{!}{\includegraphics*{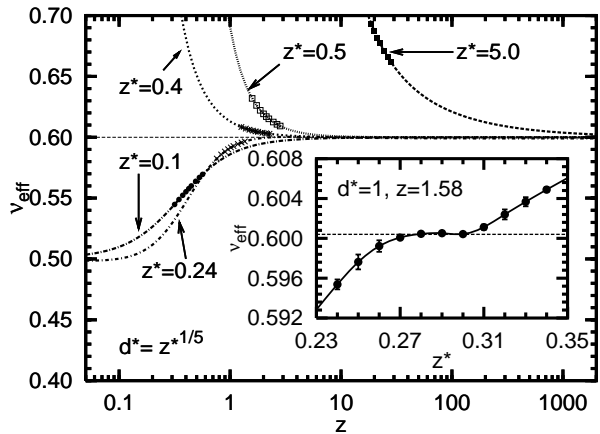}}
\caption{The exponent $\nu_\text{eff}$, in the expression
$R_\text{g} \sim N^{\nu_\text{eff}}$, as a function of $z$, and in
the inset, as a function $z^*$. \label{fig2}}
\end{figure}

Although the scaling of $R_\text{g}$ with chain length $N$ obeys a
power law only in the limit of long chains, one can make the
ansatz, $R_\text{g} \sim N^{\nu_\text{eff}}$, with
$\nu_\text{eff}$ representing an effective exponent that
approaches its critical value as $N \to \infty$. By carrying out
high precision Monte Carlo studies of lattice self avoiding walks,
Sokal and coworkers~\cite{lisokal95} have observed that
$\nu_\text{eff}$ approaches its asymptotic value \textit{from
above}, in contrast to modern two-parameter theories for the
continuum Edwards model that predict that the asymptotic value is
reached \textit{from below}. This has prompted them to suggest
that these theories do not correspond to good solvents in general,
but instead to an infinitesimal region just above the theta
temperature. Sch\"afer and co-workers~\cite{grassberger97,schafer}
have shown, by developing an alternative renormalization group
method, and by Monte Carlo simulations, that both approaches to
the asymptotic limit are not incompatible due to the existence of
a two branched structure to the solution of the excluded volume
problem. Their results indicate that the excluded volume limit is
approached on two distinct branches, depending on the whether $z^*
< z^*_\text{f}$ (the weak-coupling branch), or $z^* >
z^*_\text{f}$ (the strong-coupling branch), where $z^*_\text{f}$
denotes the \textit{fixed point} for the parameter $z^*$. While
the existence of a dual branched structure has only recently been
elucidated, the various curves in Fig.~\ref{fig2}, for
$\nu_\text{eff}$ as a function of $z$, show that Brownian dynamics
simulations readily reveal the presence of this structure. Indeed,
as will be discussed below, a unique way can be developed by which
the distinctive structure of the solution may be exploited to find
both the fixed point and the critical exponent. The curves in
Fig.~\ref{fig2} have been obtained (i) by fitting
$\alpha^2_\text{g}$ versus $z$ data, for $z^* < z^*_\text{f}$,
with $ \alpha^2_\text{g} = \left( 1 + a \, z + b \, z^2 \right)^{2
\nu -1}$; (ii) by fitting the data for $z^*
> z^*_\text{f}$ with $ \alpha^2_\text{g} = a z^{4\nu - 2} \left[ 1
- b \, z^{- |m|}\left( 1 + c/z^{-|1+m|} \right) \right]$; and
(iii) by using the fact that $\nu_\text{eff}= 0.5 + 0.25 (\partial
\ln \alpha^2_\text{g} /
\partial \ln z)$. In both the curve-fitting expressions used for
parameterizing the data (whose forms have been suggested earlier
by~\textcite{grassberger97}), the value of $\nu$ has been set
equal to 0.6.

On the weak-coupling branch (strong-coupling branch),
$\nu_\text{eff}$ first increases
(decreases) rapidly before approaching the asymptotic value very
gradually. We consequently expect a plot of $\nu_\text{eff}$ vs
$z^*$, for constant values of $d^*$ and $z$, to show a point
of inflection at the fixed point. This is
clearly seen in the inset of Fig.~\ref{fig2}, which describes the change in
$\nu_\text{eff}$ as a function of $z^*$, for a range of values of
$z^*$ spanning the fixed point. The figure suggests that
the fixed point lies in the range $0.28 \le z^*_\text{f} \le 0.3$, and
that the value of $\nu_\text{eff}$ in this range is $0.6004 \pm 0.0002$,
which is expected to be close to the critical value,
since asymptotic behavior is attained at relatively
small values of $N$ as $z^* \to z^*_\text{f}$. Sch\"afer and
co-workers~\cite{grassberger97,schafer} have estimated
$z^*_\text{f} = 0.364$, while recently \textcite{pra02}
has shown in the Gaussian approximation that, $z^*_\text{f}
\approx 0.35$. It is an exciting prospect to see if this
two branched structure persists into the non-equilibrium
regime---an issue that is easily examinable within the framework
introduced in this work.

We gratefully acknowledge the Victorian Partnership for Advanced
Computing (VPAC) for a grant under the Expertise program, and both
VPAC and the Australian Partnership for Advanced Computing for the
use of their computational facilities.

\bibliography{mybib}

\begin{thebibliography}{18}
\expandafter\ifx\csname natexlab\endcsname\relax\def\natexlab#1{#1}\fi
\expandafter\ifx\csname bibnamefont\endcsname\relax
  \def\bibnamefont#1{#1}\fi
\expandafter\ifx\csname bibfnamefont\endcsname\relax
  \def\bibfnamefont#1{#1}\fi
\expandafter\ifx\csname citenamefont\endcsname\relax
  \def\citenamefont#1{#1}\fi
\expandafter\ifx\csname url\endcsname\relax
  \def\url#1{\texttt{#1}}\fi
\expandafter\ifx\csname urlprefix\endcsname\relax\def\urlprefix{URL }\fi
\providecommand{\bibinfo}[2]{#2}
\providecommand{\eprint}[2][]{\url{#2}}

\bibitem[{\citenamefont{Noda et~al.}(1968)\citenamefont{Noda, Yamada, and
  Nagasawa}}]{nodetal}
\bibinfo{author}{\bibfnamefont{I.}~\bibnamefont{Noda}},
  \bibinfo{author}{\bibfnamefont{Y.}~\bibnamefont{Yamada}}, \bibnamefont{and}
  \bibinfo{author}{\bibfnamefont{M.}~\bibnamefont{Nagasawa}},
  \bibinfo{journal}{J. Phys. Chem.} \textbf{\bibinfo{volume}{72}},
  \bibinfo{pages}{2890} (\bibinfo{year}{1968}).

\bibitem[{\citenamefont{Solomon and Muller}(1996)}]{solmul96}
\bibinfo{author}{\bibfnamefont{M.~J.} \bibnamefont{Solomon}} \bibnamefont{and}
  \bibinfo{author}{\bibfnamefont{S.~J.} \bibnamefont{Muller}},
  \bibinfo{journal}{J. Rheol.} \textbf{\bibinfo{volume}{40}},
  \bibinfo{pages}{837} (\bibinfo{year}{1996}).

\bibitem[{\citenamefont{Sridhar et~al.}(2000)\citenamefont{Sridhar, Nguyen, and
  Fuller}}]{sridharetal00}
\bibinfo{author}{\bibfnamefont{T.}~\bibnamefont{Sridhar}},
  \bibinfo{author}{\bibfnamefont{D.~A.} \bibnamefont{Nguyen}},
  \bibnamefont{and} \bibinfo{author}{\bibfnamefont{G.~G.}
  \bibnamefont{Fuller}}, \bibinfo{journal}{J. Non-Newtonian Fluid Mech.}
  \textbf{\bibinfo{volume}{90}}, \bibinfo{pages}{299} (\bibinfo{year}{2000}).

\bibitem[{\citenamefont{Sch{\"{a}}fer}(1999)}]{schafer}
\bibinfo{author}{\bibfnamefont{L.}~\bibnamefont{Sch{\"{a}}fer}},
  \emph{\bibinfo{title}{Excluded Volume Effects in Polymer Solutions}}
  (\bibinfo{publisher}{Springer-Verlag}, \bibinfo{address}{Berlin},
  \bibinfo{year}{1999}).

\bibitem[{\citenamefont{Hayward and Graessley}(1999)}]{haygra99}
\bibinfo{author}{\bibfnamefont{R.~C.} \bibnamefont{Hayward}} \bibnamefont{and}
  \bibinfo{author}{\bibfnamefont{W.~W.} \bibnamefont{Graessley}},
  \bibinfo{journal}{Macromolecules} \textbf{\bibinfo{volume}{32}},
  \bibinfo{pages}{3502} (\bibinfo{year}{1999}).

\bibitem[{\citenamefont{Miyaki and Fujita}(1981)}]{miyaki81}
\bibinfo{author}{\bibfnamefont{Y.}~\bibnamefont{Miyaki}} \bibnamefont{and}
  \bibinfo{author}{\bibfnamefont{H.}~\bibnamefont{Fujita}},
  \bibinfo{journal}{Macromolecules} \textbf{\bibinfo{volume}{14}},
  \bibinfo{pages}{742} (\bibinfo{year}{1981}).

\bibitem[{\citenamefont{Bercea et~al.}(1999)\citenamefont{Bercea, Ioan, Ioan,
  Simionescu, and Simionescu}}]{berceaetal99}
\bibinfo{author}{\bibfnamefont{M.}~\bibnamefont{Bercea}},
  \bibinfo{author}{\bibfnamefont{C.}~\bibnamefont{Ioan}},
  \bibinfo{author}{\bibfnamefont{S.}~\bibnamefont{Ioan}},
  \bibinfo{author}{\bibfnamefont{B.~C.} \bibnamefont{Simionescu}},
  \bibnamefont{and} \bibinfo{author}{\bibfnamefont{C.~I.}
  \bibnamefont{Simionescu}}, \bibinfo{journal}{Prog. Polym. Sci.}
  \textbf{\bibinfo{volume}{24}}, \bibinfo{pages}{379} (\bibinfo{year}{1999}).

\bibitem[{\citenamefont{Doi and Edwards}(1986)}]{doiedw}
\bibinfo{author}{\bibfnamefont{M.}~\bibnamefont{Doi}} \bibnamefont{and}
  \bibinfo{author}{\bibfnamefont{S.~F.} \bibnamefont{Edwards}},
  \emph{\bibinfo{title}{The Theory of Polymer Dynamics}}
  (\bibinfo{publisher}{Oxford University Press}, \bibinfo{address}{Oxford},
  \bibinfo{year}{1986}).

\bibitem[{\citenamefont{Li et~al.}(1995)\citenamefont{Li, Madras, and
  Sokal}}]{lisokal95}
\bibinfo{author}{\bibfnamefont{B.}~\bibnamefont{Li}},
  \bibinfo{author}{\bibfnamefont{N.}~\bibnamefont{Madras}}, \bibnamefont{and}
  \bibinfo{author}{\bibfnamefont{A.~D.} \bibnamefont{Sokal}},
  \bibinfo{journal}{J. Stat. Phys.} \textbf{\bibinfo{volume}{80}},
  \bibinfo{pages}{661} (\bibinfo{year}{1995}).

\bibitem[{\citenamefont{Grassberger et~al.}(1997)\citenamefont{Grassberger,
  Sutter, and Sch\"afer}}]{grassberger97}
\bibinfo{author}{\bibfnamefont{P.}~\bibnamefont{Grassberger}},
  \bibinfo{author}{\bibfnamefont{P.}~\bibnamefont{Sutter}}, \bibnamefont{and}
  \bibinfo{author}{\bibfnamefont{L.}~\bibnamefont{Sch\"afer}},
  \bibinfo{journal}{J. Phys. A: Math. Gen.} \textbf{\bibinfo{volume}{30}},
  \bibinfo{pages}{7039} (\bibinfo{year}{1997}).

\bibitem[{\citenamefont{Graessley et~al.}(1999)\citenamefont{Graessley,
  Hayward, and Grest}}]{greetal99}
\bibinfo{author}{\bibfnamefont{W.~W.} \bibnamefont{Graessley}},
  \bibinfo{author}{\bibfnamefont{R.~C.} \bibnamefont{Hayward}},
  \bibnamefont{and} \bibinfo{author}{\bibfnamefont{G.~S.} \bibnamefont{Grest}},
  \bibinfo{journal}{Macromolecules} \textbf{\bibinfo{volume}{32}},
  \bibinfo{pages}{3510} (\bibinfo{year}{1999}).

\bibitem[{\citenamefont{{\"{O}}ttinger}(1989)}]{ott89c}
\bibinfo{author}{\bibfnamefont{H.~C.} \bibnamefont{{\"{O}}ttinger}},
  \bibinfo{journal}{Phys. Rev. A} \textbf{\bibinfo{volume}{40}},
  \bibinfo{pages}{2664} (\bibinfo{year}{1989}).

\bibitem[{\citenamefont{Zylka and {\"{O}}ttinger}(1991)}]{zylott91}
\bibinfo{author}{\bibfnamefont{W.}~\bibnamefont{Zylka}} \bibnamefont{and}
  \bibinfo{author}{\bibfnamefont{H.~C.} \bibnamefont{{\"{O}}ttinger}},
  \bibinfo{journal}{Macromolecules} \textbf{\bibinfo{volume}{24}},
  \bibinfo{pages}{484} (\bibinfo{year}{1991}).

\bibitem[{\citenamefont{Prakash}(2001{\natexlab{a}})}]{pra01a}
\bibinfo{author}{\bibfnamefont{J.~R.} \bibnamefont{Prakash}},
  \bibinfo{journal}{Macromolecules} \textbf{\bibinfo{volume}{34}},
  \bibinfo{pages}{3396} (\bibinfo{year}{2001}{\natexlab{a}}).

\bibitem[{\citenamefont{Prakash}(2002)}]{pra02}
\bibinfo{author}{\bibfnamefont{J.~R.} \bibnamefont{Prakash}},
  \bibinfo{journal}{J. Rheol.} \textbf{\bibinfo{volume}{46}},
  \bibinfo{pages}{1353} (\bibinfo{year}{2002}).

\bibitem[{\citenamefont{{\"{O}}ttinger}(1996)}]{ottbk}
\bibinfo{author}{\bibfnamefont{H.~C.} \bibnamefont{{\"{O}}ttinger}},
  \emph{\bibinfo{title}{Stochastic Processes in Polymeric Fluids}}
  (\bibinfo{publisher}{Springer-Verlag}, \bibinfo{year}{1996}).

\bibitem[{\citenamefont{Prakash}(2001{\natexlab{b}})}]{pra01b}
\bibinfo{author}{\bibfnamefont{J.~R.} \bibnamefont{Prakash}},
  \bibinfo{journal}{Chem. Eng. Sci.} \textbf{\bibinfo{volume}{56}},
  \bibinfo{pages}{5555} (\bibinfo{year}{2001}{\natexlab{b}}).

\bibitem[{\citenamefont{Kumar and Prakash}()}]{satpra02}
\bibinfo{author}{\bibfnamefont{K.~S.} \bibnamefont{Kumar}} \bibnamefont{and}
  \bibinfo{author}{\bibfnamefont{J.~R.} \bibnamefont{Prakash}},
  \bibinfo{note}{to be published}.

\end{thebibliography}
\end{document}